\pdfoutput=1
\documentclass[twocolumn,prb,showpacs,aps,superscriptaddress]{revtex4-1}

\usepackage[english]{babel}
\usepackage[ansinew]{inputenc}
\usepackage{times}
\usepackage{graphicx}
\usepackage{graphics}
\usepackage{amsmath}
\usepackage{amsfonts}
\usepackage{amssymb}
\usepackage{amsbsy}
\usepackage{dsfont}
\usepackage{epstopdf}
\usepackage{makeidx}
\usepackage{color} 
\usepackage{pgf}
\usepackage{bm}
\usepackage{tikz} 
\usepackage{graphicx}  % needed for figures
\usepackage{dcolumn}   % needed for some tables
\usepackage{bm}        % for math
\usepackage{amssymb}   % for math

\newcommand{\be}{\begin{equation}}
\newcommand{\ee}{\end{equation}}
\newcommand{\bea}{\begin{eqnarray}}
\newcommand{\eea}{\end{eqnarray}}

\makeindex
\begin{document}

\title{Quantum-Accurate Molecular Dynamics Potential for Tungsten}

\author{M. A. Wood}
\affiliation{Center for Computing Research, Sandia National Laboratories, Albuquerque, New Mexico 87185, USA}
\author{A. P. Thompson}
\affiliation{Center for Computing Research, Sandia National Laboratories, Albuquerque, New Mexico 87185, USA}

\date{\today}

\begin{abstract}
The purpose of this short contribution is to report on the development of a Spectral Neighbor Analysis Potential (SNAP) for tungsten. 
We have focused on the characterization of elastic and defect properties of the pure material in order to support molecular dynamics simulations of plasma-facing materials in fusion reactors. 
A parallel genetic algorithm approach was used to efficiently search for fitting parameters optimized against a large number of objective functions.  
In addition, we have shown that this many-body tungsten potential can be used in conjunction with a simple helium pair potential\cite{Juslin2013} to produce accurate defect formation energies for the W-He binary system. 
\end{abstract}

\pacs{}

\maketitle
\section{\label{sec:level1}Introduction}
Advancements in the field of classical molecular dynamics simulation can be classified into three general categories; i) those that increase the length scales of accessible problems, ii) extensions of the timescales accessible by modern software packages and lastly, iii) improvements to the accuracy, quality, or realism of predictions made by MD. 
In other words these three areas, in one way or another, correspond to the computing hardware, software and fidelity of the approximations that are used to run the simulation.
But while these categories are distinct, it would be wrong to assume that there is no interdependence between them. In fact, any one of these areas may be pushed forward by an advancement in any of the other.

The work presented here will fall into the third general category of improved fidelity, where we are developing interatomic potentials (IAP) that, as we will show here, approach the accuracy of Density Functional Theory (DFT) while preserving the scalability of classical MD simulations.
Of course the distillation of a truly many-body energy functional from DFT into energies and forces that are written on a per-atom basis comes with certain approximations.
Classical MD necessarily uses purely local information to calculate the force on each atom and the total energy of the system.  Without this near-sightedness, 
it would not be possible to use the scalable parallel algorithms that enable simulations of millions of atoms on large distributed computer systems.
Even with this short-ranged approximation, many IAP have been successful in including multi-body effects through the addition of higher-order terms in the total energy that take into account many-body aspects of the local environment such as bond angles, dihedrals, coordination dependence, or even more generally bond-order dependent terms.
While the properties of materials are best determined from an \emph{ab initio} level of theory, the study of defects and failure mechanisms necessitates a length scale that is pushed beyond what is tractable at this level of theory.
In this work, we are concerned with the defect interactions and failure modes of plasma-facing components, namely tungsten, in fusion reactors where surface defect features range in size from sub-nanometers to millimeters.
At the smallest length scales, an accurate description of a wide range of these defects is challenging because the local environment of defect cores is unlike the crystalline matrix.
Therefore, the IAP in MD must be able to simultaneously capture the potential energy surface of atoms in either of these bonding configurations. Therefore, successful models must in some way represent multi-body interactions.
\section{\label{sec:level1}SNAP Formalism}
A recent trend in molecular dynamics is to integrate the advancements from the fields of machine-learning and neural networks in order to produce more accurate potential energy surfaces. 
In essence, this class of IAP dispense with a fixed functional form in favor of more general statistical relationships that are adjusted under supervised training to reproduce a large set of training data.  
This supervised training data is supplied by the user and consists of a collection of high-accuracy results from \emph{ab initio} calculations. 
The statistical relationships uses a set of descriptors to map atomic configurations to specific values of energy and force.
The main differences between potentials lie in the set of descriptors that are used\cite{Bartok2010}.
The set of descriptors used in the machine-learning process must be physically motivated, because the potential must be able to distinguish unique atomic environments, while at the same time satisfy basic physical principles, such as permutation, translational and rotation invariance.\cite{Ferre2015}

The SNAP interatomic potential falls under this umbrella of machine-learned potentials in that it uses a set of descriptors of the local environment around an atom in order to calculate the per-atom energy function.
This is in direct contrast to most IAP that use a fixed functional form with a single or few dependent variables (such as distance, bond order or angle) to calculate the interactions of atoms.
A key advantage of a flexible IAP such as SNAP is that its accuracy is directly dependent on the number of bispectrum components\cite{BartokThesis} used in the atomic energy expansion.  
Hence it is possible to systematically improve the accuracy of the potential, at the price of increased computational cost.
%Therefore the SNAP formalism shares a common feature with GAP potentials which is that have a tunable accuracy during the fitting procedure through the use of the bispectrum atom descriptors.
Following the GAP formalism of Bartok et al.\cite{Bartok2010}, SNAP uses the bispectrum components of the 4D hyperspherical harmonics on the 3-sphere to characterize atomic neighborhoods.
A key difference between the GAP and SNAP formalism is that SNAP assumes a linear relationship between the atomic energy and the bispectrum components.
The details of the derivation of the SNAP energy function are described elsewhere \cite{Thompson2015}. 
Here we provide a brief summary. 
The potential energy of a configuration of atoms with is written as the sum of a reference potential and the SNAP contribution
\begin{equation}
E({\bf r}^{N})=E_{ref}({\bf r}^{N})+E_{\tiny{SNAP}}({\bf r}^{N}),
\label{eq1}
\end{equation}

where $E$, $E_{Ref}$, and $E_{SNAP}$ are the total, reference, and SNAP potential energies, respectively.  
The term ${\bf r}^N$ is a vector giving the positions of the $N$ atoms in the configuration.
In the present case for tungsten, the ZBL pair potential is used to capture the short-range repulsive interactions between atomic cores. 
%[APT I am pretty sure you are not using charges.]
% and a long-range electrostatic term. 
Within the SNAP formalism, $E_{SNAP}$ is expressed as a sum over individual atom energies, each of which is 
in turn expressed as a weighted sum over its bispectrum components.
\begin{eqnarray}
E_{\tiny{SNAP}}({\bf r}^{N}) & = & \sum_{i=1}^{N}{\beta}_{0}+\sum_{k=1}^{K}{\beta_k}\large{B}_{k}^{i} \\
& = & {\beta}_{0}N+\boldsymbol\beta\cdot\sum_{i=1}^{N}\large{\bf B}^{i}
\label{eq2}
\end{eqnarray}
Similarly, the force on each atom $j$ due to the SNAP potential can be expressed as a weighted sum over the derivatives w.r.t. $j$ of the bispectrum components of each atom $i$.
\begin{equation}
{\bf F}^{j}_{\tiny{SNAP}}=-\nabla_{j}E_{\tiny{SNAP}}=-\boldsymbol\beta{\cdot}\sum_{i=1}^{N} \frac {\partial {\bf \large{B}}^{i}}{\partial {\bf r}_{j}}
\label{eq3}
\end{equation}

In this way, the total energy, forces, and also the stress tensor, can be written as linear function of quantities related to the bispectrum components of the atoms. 
Given a large set of training data, it is straightforward to construct a weighted least squares linear regression problem whose solution provides optimal choices for the coefficients $\beta_0,...,\beta_K.$  
Less straightforward is the determination of optimal values for the so-called hyper-parameters that define the regression problem. 
These include parameters that define the neighborhood of each atom, as well as parameters that control the extent to which different components of the training data are prioritized. 
The optimization of the hyperparameters is discussed further in the following section. 

%The goal of the fitting process is then to determine the weights($\beta$) that multiplies each bispectrum term in order the minimize the discrepancy between the provided DFT training data to the energies and forces predicted by SNAP.
%As we will show in this report, this linear form of Equation \ref{eq2} allows for a rapid fitting procedure and enables an efficient search for candidate potentials for a wide range of material properties. 
%

\section{\label{sec:level1}Fitting Procedure}
One of the key approximations for interatomic potentials is the ability to accurately predict the properties of materials/molecules that were not in the training data. 
This is generally referred to as the transferability of an IAP.
Normally, the end users select an IAP based on its interpreted performance for the experiments they plan to run, we as the developers of said MD potential are not completely absolved from these constraints either.
Therefore we must be aware of each step in the fitting procedure that has the ability to bias or affect the accuracy of the IAP for its intended use.
We have identified three main areas where this could occur; i) in the selection of the training set used in supervised learning, ii) the choice of the objective functions used during fit optimization and iii) interpretation of the optimized hyper-parameters as it translates to the quality of the IAP. 
The present work is focused on the accuracy of tungsten potentials for use in simulation of plasma-facing materials in a fusion reactor.
Therefore, we have chosen training data and fitting objectives that reflect properties relevant to this application.
It is true that, in its present state, the reported SNAP potential has not been rigorously tested for properties outside the training set.
However, as will be shown in the following section, the addition of specialized training data during the fitting process can correct for these unforeseen inaccuracies. 
In addition, the outputs of the fitting process provide a transparent means to compare the importance of each training configuration for reproducing the reference data.
We believe this is a particular strength of the SNAP formalism because it has the capability for end users to be better informed to the strength and weakness of the IAP.
The more aware we are of the human bias (which may never be completely removed) that is involved in generating an IAP, the better the field of MD users can be at objectively selecting and interpreting this key approximation of MD simulation results.

%%%
\begin{figure}[!t]
\includegraphics{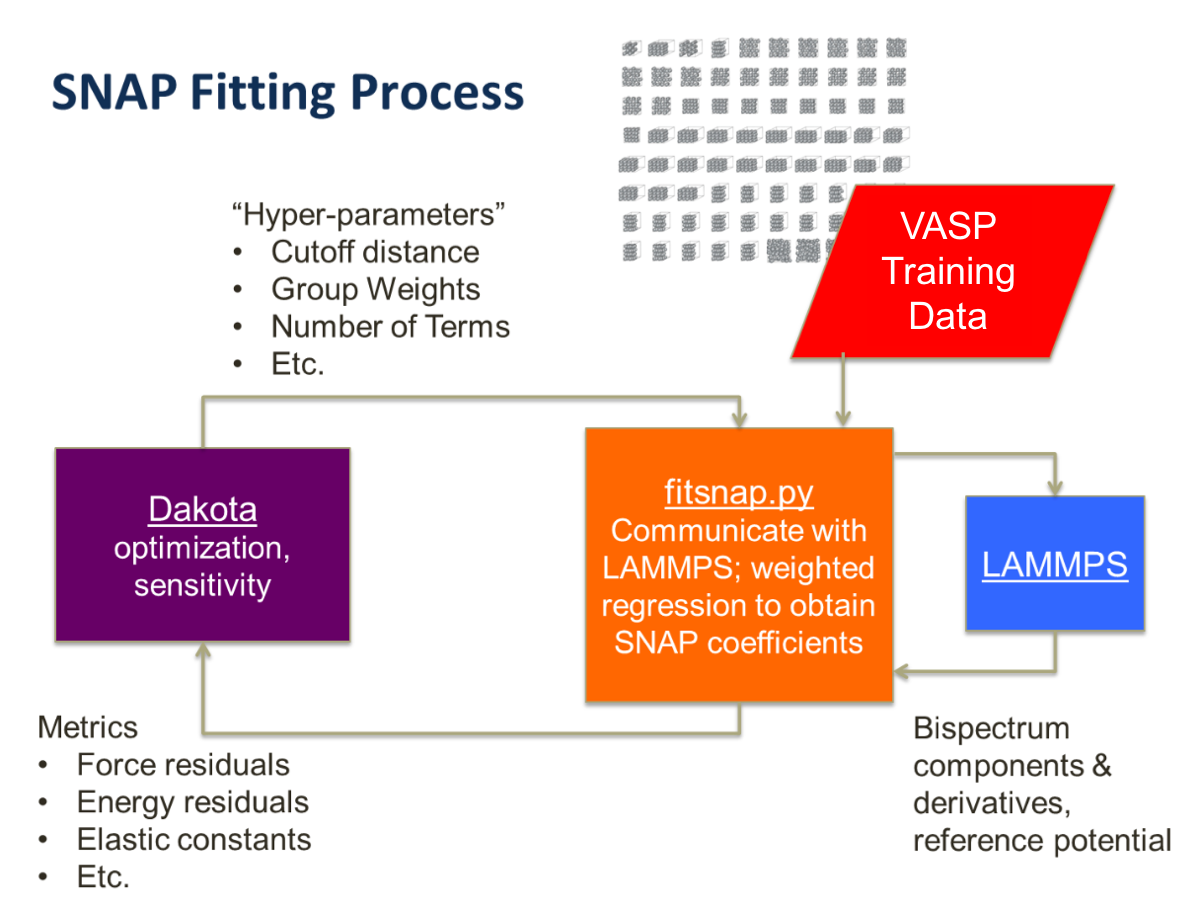}% 
\caption{\label{fig1}Schematic of the SNAP fitting procedure.}
\end{figure}
%%%
\subsection{\label{sec:level2}Selection of Training Data}
Much like the parameterization of other IAP, the goal of the fitting process is to minimize the discrepancies that occur between reference data and the predictions made by the new candidate potential.
Most of the fitting protocol to generate SNAP potentials is automated and requires little user input after the training structures are generated. 
However, the selection of reference structures is still an area where expert judgement is required, as it has a strong influence on the accuracy and transferability of the resulting potential. 
We chose to begin with the existing set of training data used to construct the GAP tungsten potential \cite{Szlachta2014}, which is publicly available online\cite{libatoms}.
The GAP training data consists of positions, total energies$(E)$, forces on each atom$(F)$ and virial stresses$(\sigma)$ for several thousand small configurations of tungsten atoms. 
We selected a subset of this training data, omitting those structures that we deemed less relevant to our application.

This addition or subtraction of training data for the needs of the potential is a unique trait of machine-learned potentials, but to date there are only empirical guidelines for how this should be managed.
It is a future direction of this work to quantify the impact of certain training data on the quality of the generated potential.
A sound understanding of the translation of the DFT potential energy surface to an IAP used in MD will help researchers more efficiently deploy the extensive computing resources needed to large training data sets.
Again, this is an avenue for future work and will not be addressed in detail here.

During the fitting routine it is advantageous to limit the number of hyperparameters in order to limit the dimensionality of the search space. 
In the case of SNAP, this means grouping training structures together in order to reduce the degeneracy of the information provided by any one training geometry.
The process by which these groups are defined is necessarily subjective.  
Our goal was to define these groups based on the differences in bonding environments or defect types.
From the GAP training potential we selected four groups pertaining to primitive cell deformations (x2000 configurations), a super-cell of BCC tungsten with a single vacancy (x420), free surfaces with various orientations (x180) and a sampling of a DFT-MD run of the BCC crystal at 300K (x60).
In contrast, omitted training data pertaining to dislocations and vacancy-dislocation binding, which was used to fit the GAP potential.
We expect that in order to accurately capture defect properties relevant to helium implantation in tungsten, a wider variety of bonding environments such as over- and under-coordinated atoms must be included in the fit. 
Therefore, three new training data groups consisting of a few hundred new geometries were added.
The isotopic compression and dilation of a BCC, FCC, A15 and simple cubic tungsten crystals form the Equation of State category shown in Table \ref{table1}.
In addition, a 54 atom supercell of tungsten was melted and a set of 27 of these liquid structures were used to for training. 
Lastly, $3\times3\times3$ BCC crystals with a pair of vacancies ranging from the nearest- to fourth nearest-neighbor positions were included in the fit. 
Each of these training geometries was calculated using the VASP\cite{VASPcode} plane wave DFT code with a $8\times8\times8$ Monkhorst Pack k-point mesh, 600eV plane wave cutoff energy, 0.1eV Gaussian smearing, PBE exchange-correlation functional and a GW pseudopotential.

\subsection{\label{sec:level2}Optimization of a SNAP Potential}
Once the components of the training data were assembled, the Single Objective Genetic Algorithm (SOGA) capability within the optimization software DAKOTA\cite{Dakota2006} is used in order to search for optimal choices of the hyperparameters. 
These hyperparameters are the radial cutoff used to define the SNAP neighborhood of each atom, and the weights used in formulating the least squares linear regression problem for the energy, force, and virial stress data of each training group.
%Using Figure \ref{fig1} as a visual guide, these search parameters are denoted as 'Hyper-parameters' in the main loop of the SNAP parameterization loop.
Each of these regression weights is allowed to vary from $10^0$ to $10^{4}$ in order to properly explore its importance to the overall fit quality. 
It is worth mentioning that the absolute values of the regression weights are arbitrary and only carry meaning when compared to other regression weights.
The last fitting parameter is the radial cutoff which is allowed to vary between 4{\AA} and 5{\AA}, excluding or including the third nearest neighbor interactions at either limit.
For a given choice of hyperparameters, the weighted least squares linear regression problem is formulated and then solved using QR decomposition, yielding optimal values for the vector $\boldsymbol\beta$, representing a particular candidate potential.

Of equal importance to the input training data is the set objective functions that  SOGA uses to evaluate the candidate potentials for evolutionary fitness. 
These objectives will determine where the resultant IAP is certain to provide accurate results and should be chosen such that they also span a range of distinct material properties in order to provide some sense of transferability.
The SOGA objective function is composed as the sum of eleven distinct error metrics.
These include the relative error in the following properties:  lattice parameter, three elastic constants of the BCC phase, five basic defect formation energies.
Also included is the absolute error in energies and forces summed over all training data. 
The error metrics are either obtained directly from fitsnap.py (see Figure \ref{fig1}) or though a LAMMPS simulation of elastic constants or defect energies run using the current candidate potential.
After three-hundred candidates have been evaluated, the genetic algorithm will take the highest ranking fits and perform breeding and mutation steps to initialize the next generation of candidates.
In this gradient free genetic algorithm the search could in principle continue indefinitely.  
However, we saw steady improvement in the overall accuracy metric (sum of all errors) for the first ten generations, after which no significant improvement was observed.
%%%
\begin{table}[!t]
\caption{Resulting fitting weights on the provided DFT training data for the SNAP potential. Values in each column for $E$, $F$ and $\sigma$ are calculated using Equation \ref{eq5}-\ref{eq7}, respectively}
\begin{ruledtabular}
\begin{tabular}{lccccc}
&&$w^{total}_{E}$&$w^{total}_{F}$&$w^{total}_{\sigma}$\\
Category&$N_{config}$&(eV/atom)&(eV/{\AA})&(bar)\\
\hline
\\
Elastic Deform\footnotemark[1]&2000&$1.0\cdot10^{7}$&-&$5.2\cdot10^{4}$\\
Single Vacancy\footnotemark[1]&420&$3.2\cdot10^{4}$&$4.8\cdot10^{6}$&-\\
Free Surface\footnotemark[1]&180&$4.3\cdot10^{3}$&$3.9\cdot10^{7}$&-\\
DFT-MD, 300K\footnotemark[1]&60&$6.3\cdot10^{2}$&$1.1\cdot10^{7}$&-\\
Equation of State\footnotemark[2]&168&$3.7\cdot10^{2}$&$7.2\cdot10^{3}$&-\\
Liquids\footnotemark[2]&27&$8.2\cdot10^{2}$&$2.9\cdot10^{7}$&-\\
Multiple Vacancies\footnotemark[2]&20&$1.0\cdot10^{5}$&$2.3\cdot10^{4}$&-\\
\end{tabular}
\end{ruledtabular}
\footnotetext[1]{From Szlachta, Reference [\onlinecite{Szlachta2014}]}
\footnotetext[2]{Current Work}
\label{table1}
\end{table}
%%%
%Elastic Deform\footnotemark[1]&2000&\bf{5105}&-&4.3\\
%Single Vacancy\footnotemark[1]&420&23.9&5927&-\\
%Free Surface\footnotemark[1]&180&77.2&65.6&-\\
%DFT-MD, 300K\footnotemark[1]&60&10.5&458.2&-\\
%Equation of State\footnotemark[2]&168&2.21&1.60&-\\
%Liquids\footnotemark[2]&27&30.2&\bf{9338}&-\\
%Multiple Vacancies\footnotemark[2]&20&5037&3.79&-\\

\subsection{\label{sec:level2}Interpretation of Fitting Parameters}
After an optimized potential is generated from the previous steps, a natural question to ask is how much did each group of training data influence the overall objective function?
Table \ref{table1} displays the resultant fitting parameters for the best performing SNAP result using the SOGA algorithm.
In this table it is useful to compare the effective weights of each category for a given property ($E$, $F$ or $\sigma$) and see which training geometries dominate the resultant fit($R_{cut}=4.734${\AA}).
These effective weights show the relative importance of different types of training data in the final fit and are given by the
following formulas

\begin{eqnarray}
\label{eq5}
w^{total}_{E}=\phi^{j}_{E}\cdot N_{config}\\
\label{eq6}
w^{total}_{F}=\phi^{j}_{F}\cdot\sum_{i=1}^{N_{config}}3\cdot N^{i}_{atoms}\\
\label{eq7}
w^{total}_{\sigma}=\phi^{j}_{\sigma}\cdot (6\cdot N_{config})
\end{eqnarray}
In Equations \ref{eq5}-\ref{eq7}, $\phi^{j}_\alpha$ is the weight assigned to group $j$ for property $\alpha$.  
These are the weights adjusted during the DAKOTA optimization.
For each training configuration, there is just one energy term and six components of the stress tensor.
Meanwhile, there are three forces per atom for each configuration which is shown in Equation \ref{eq6} as an explicit sum over the number of configurations in group $j$.
The effective weights, not group weights, should be used to judge the relative importance of each of type of training data.
The largest effective weight for energies belongs to the set of elastic deformation training data while the free surface and liquid configurations were roughly equivalent in terms of their importance in producing accurate forces. 
Including the liquid structures was precisely for this reason, though the result was not guaranteed.
A liquid samples a number of unique under and over coordinated bonding environments that we believe translates to better predictions of defect properties in tungsten where atoms are forced into energetically unfavorable configurations. 
It is not surprising then to see the two most 'abnormal' bonding environments, free surface and liquid configurations, to dominate the fit for producing accurate forces.
\section{\label{sec:level1}Results}
Here we draw comparisons to a handful of well known W-potentials that span a variety of functional forms and computational costs, this includes the GAP potential that shares training data with the current SNAP potential.  

It is common for any IAP to test the accuracy of the available structural and mechanical properties for the material of interest, these are most commonly the lattice parameter and elastic constants of the stable phases.
For tungsten, this is not a complicated task since the $\alpha$ phase (BCC) is the only stable structure of the pure phase so there are far fewer target structural properties to meet.
The A15 phase is predicted to be metastable below the melting temperature, but is not included as a fitting objective of the optimization process, we have however confirmed that the A15 phase is less stable than BCC tungsten at zero kelvin.
The collection of predicted elastics constants as they compare to the DFT data used as the training reference as well as an EAM, MEAM and GAP potential are shown in Table \ref{table2}.
Excluding the DFT data, all values in Table \ref{table2} and \ref{table3} were calculated by the current authors using the LAMMPS MD package which contains all of these potentials.
Where available our predictions are compared to reported results in the original published work for each potential all of which agree within the numerical precision that is reported in our tables.
The fact that each of these potentials have been run by the current authors using a single MD engine will become very important when timing data is presented later in this section. 
Most of the potentials show good accuracy for the lattice constant and elastic constants of the $\alpha$ phase, with the maximum discrepancy with respect to the DFT data being a few GPa or single digit percent error. The one exception is the MEAM potential, which is significantly less accurate for the elastic constants.
This result is encouraging for the SNAP potential presented here, the extremely low discrepancy to DFT indicates that the SNAP formalism can easily capture atomic interactions in a crystalline system.

%%%
\begin{table}[t]
\caption{\label{tab:table4}Structural and mechanical properties of tungsten predicted from common interatomic potentials}
\begin{ruledtabular}
\begin{tabular}{lccccc}
 & DFT\footnote{Derlet, Reference[\onlinecite{Derlet2007}]}&EAM\footnote{Juslin, Reference [\onlinecite{Juslin2013}]}&MEAM\footnote{Scheiber, Reference [\onlinecite{Scheiber2016}]}&GAP\footnote{Szlachta, Reference [\onlinecite{Szlachta2014}]}&SNAP\footnote{Current Work}\\
\hline
$a_{0}$ ({\AA})&3.1803&3.165&3.164&3.1803&3.1805\\
$C_{11}$ (GPa)&517&517&533&518&518\\
$C_{12}$ (GPa)&198&200&205&198&195\\
$C_{44}$ (GPa)&142&156&163&143&144\\
\end{tabular}
\end{ruledtabular}
\label{table2}
\end{table}
%%%

On the other hand, the accuracy of predicted defect formation energies shows much greater variability over the different tungsten potentials.
The top section of Table \ref{table3} shows the defect formation energies for a pure tungsten system which includes the self-interstitial at the tetrahedral and octahedral sites, dumbbell defects in the [110] and [111] directions, single vacancies and lastly the binding energy between two vacancies in the nearest-neighbor positions.
Each of these defect formation energies, except for the divacancy binding energy, were used as a fitting objective for optimizing SNAP in DAKOTA.  In other words, in addition to reproducing the DFT forces and energies, the SNAP parameterization process attempts to minimizes these defect formation errors.
Hence, it is not unexpected that the current SNAP potential outperforms all of the previous tungsten potentials.
It is impressive that the SNAP potential predicts a repulsive divacancy binding energy at the nearest-neighbor (NN) position, while both both the EAM and MEAM potentials
incorrectly predict a weak attractive interaction. SNAP was not explicitly trained for this property.  In the high temperature environment of a fusion reactor, this property
plays a key role in He bubble formation. 
To further test this potential for material properties outside the training set that are relevant to the fusion application, 
we have chosen to incorporate the He/He and He/W  pair potentials developed by Juslin and Wirth for use with the 
EAM tungsten potential. \cite{Juslin2013}  We have calculated defect formation energies of helium defects 
within the tungsten matrix. The results are displayed in the lower section of Table \ref{table3}.

%%%
\begin{table}[!t]
\caption{\label{tab:table4}Defect formation energies for BCC tungsten (eV)}
\begin{ruledtabular}
\begin{tabular}{lcccccc}
&DFT\footnote{W: Derlet, Reference [\onlinecite{Derlet2007}], He: Becquart, Reference [\onlinecite{Becquart2007}]}&EAM\footnote{Juslin, Reference [\onlinecite{Juslin2013}]}&MEAM\footnote{Scheiber, Reference [\onlinecite{Scheiber2016}]}&GAP\footnote{Szlachta, Reference [\onlinecite{Szlachta2014}]}&SNAP\footnote{Current Work}\\
\hline
$E_{W-Tetra}$&11.1 &10.4&11.3&12.8&11.1\\
$E_{W-Octa}$&11.7 &10.4&13.6&13.1&11.5\\
$E_{W-[110]d}$&9.8 &10.3&9.1&11.8&9.8\\
$E_{W-[111]d}$&9.6&9.8&10.4&11.1&9.7\\
$E_{W-vacancy}$&3.3 &3.7&4.0&3.3&3.2\\
$E_{W-divacancy}\footnote{Binding energy at nearest neighbor positions}$&0.1 &-0.4 &-0.2 &0.4 &0.1\\
\hline
$E_{He-Tetra}$&6.2 &6.2&6.4&6.3&6.4\\
$E_{He-Octa}$&6.4 &6.2&6.4&6.3&6.3\\
$E_{He-Subs}$&4.7 &4.8&5.3&4.3&4.3\\
$E_{2He-Tetra}\footnote{Binding energy at adjacent tetrahedral sites}$&1.0 &0.9&0.6&0.8&1.0\\
\end{tabular}
\end{ruledtabular}
\label{table3}
\end{table}
%%%

The defect types calculated are the insertion of a single He into the tetrahedral or octahedral sites of BCC W, a substitution of He for a single W atom and the binding energy of two He atoms in adjacent tetrahedral sites of BCC tungsten.
Comparing the relative error of each potential, the SNAP potential predicts formation energies within a few percent ($<3.5\%$) of the DFT reference for all defects, except the He substitution, which is $~9\%$ lower than the expected value of 4.70~eV.

\begin{figure}[t]
\includegraphics{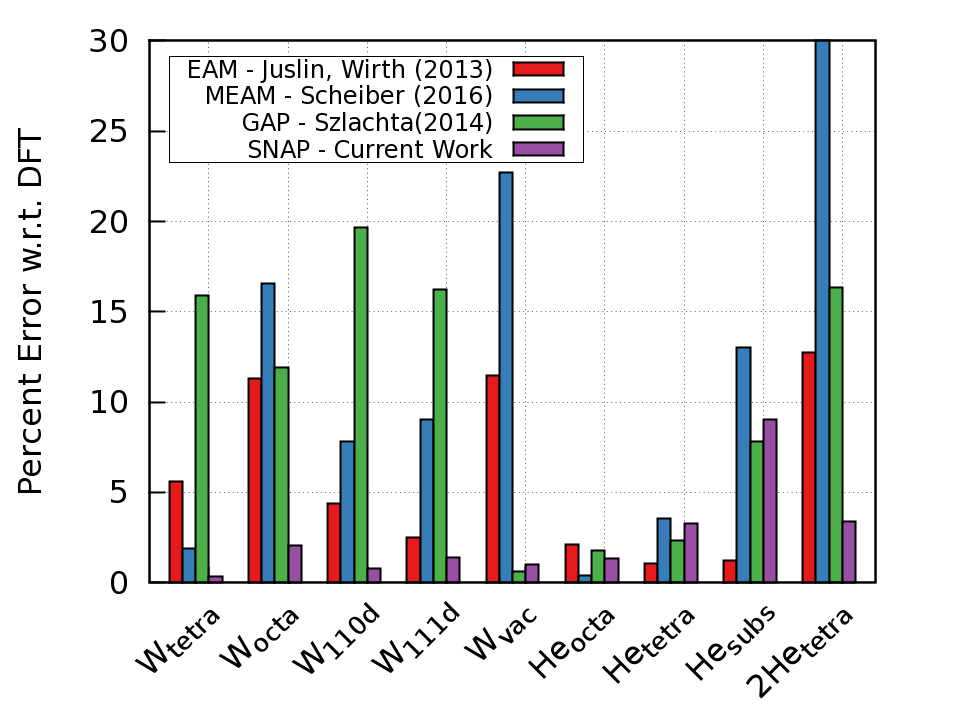}% 
\caption{\label{fig2}Collection of W-defect formation energies for commonly used interatomic potentials. The raw data shown here is also contained in Table \ref{table3}, DFT reference data taken from Derlet \emph{et. al.} (2007) and Becquart \emph{et. al.} (2007) for W and He defects, respectively.}
\end{figure}

The relative errors across all potentials tested here are visually displayed in Figure \ref{fig2}.
It is interesting to note the differences in the GAP and SNAP potentials.  These two IAP share much of the same training data.
However, the GAP training data did not include any point defect structures, other than the single vacancy structure.
Similarly, the SNAP training data also did not include any point defect structures, other than the divacancy structures. 
However, by including defect formation energies as objectives, the DAKOTA optimization guided the SNAP potential 
towards high accuracy for the tungsten defect structures.
Interestingly, these two machine-learned potentials show great flexibility in which material properties can be accurately captured.
This is a feature of the atomic descriptors that underline the potential energy surface of both.
Again we should emphasize that the SNAP potential here was trained against material properties that are relevant to the simulation of tungsten as a diverter material in fusion reactors.
The high temperature helium environment guided the choice of these defects as objective functions.
It is a difficult to directly assess the loss of transferability of this SNAP IAP because the number of configurations and material properties outside the training set vastly exceeds those that are included.
As a new research direction, we are currently exploring the strategy of sampling MD trajectories generated using the SNAP potential and 
testing their accuracy against DFT.
For now, this new SNAP potential has been shown to accurately reproduce elastic constants and defect formation energies to high accuracy.
These properties are essential for the prediction of He bubble formation in plasma-facing materials.
There is room for improvement in the accuracy of the He defects. These are the largest errors predicted by the SNAP potential.
We see the inclusion of W-He and He-He interactions into the SNAP formalism as the next stages of this work, and 
progress is currently being made in this avenue. 

%%%
%%%

Before closing this section it is worth addressing the problem of computational cost of each of the tungsten potentials that are highlighted here.
It is true that the added accuracy of GAP or SNAP comes with a significant computational cost with respect to EAM or MEAM.
It is important to provide realistic timing data so that users are well informed of the strengths and weaknesses of each potential.
Each of these potentials is implemented into the LAMMPS MD software package\cite{Plimpton1995}, 
which is freely available online \cite{lammpsweb}. 
The EAM, MEAM, and SNAP potentials can be invoked within any standard build of LAMMPS, provided it has been compiled with the
corresponding optional package, {\tt MANYBODY}, {\tt MEAM}, {\tt SNAP}, respectively.  The GAP potential requires
compiling LAMMPS with the {\tt USER-QUIP} optional package, which provides a library interface to the QUIP software package.\cite{QUIP} 
Further documentation, input files and simple test problems for this SNAP potential are now included in the LAMMPS distrbution.
More technical details about this potential can be found online \cite{lammpssnap}.

\begin{figure}[!t]
\includegraphics{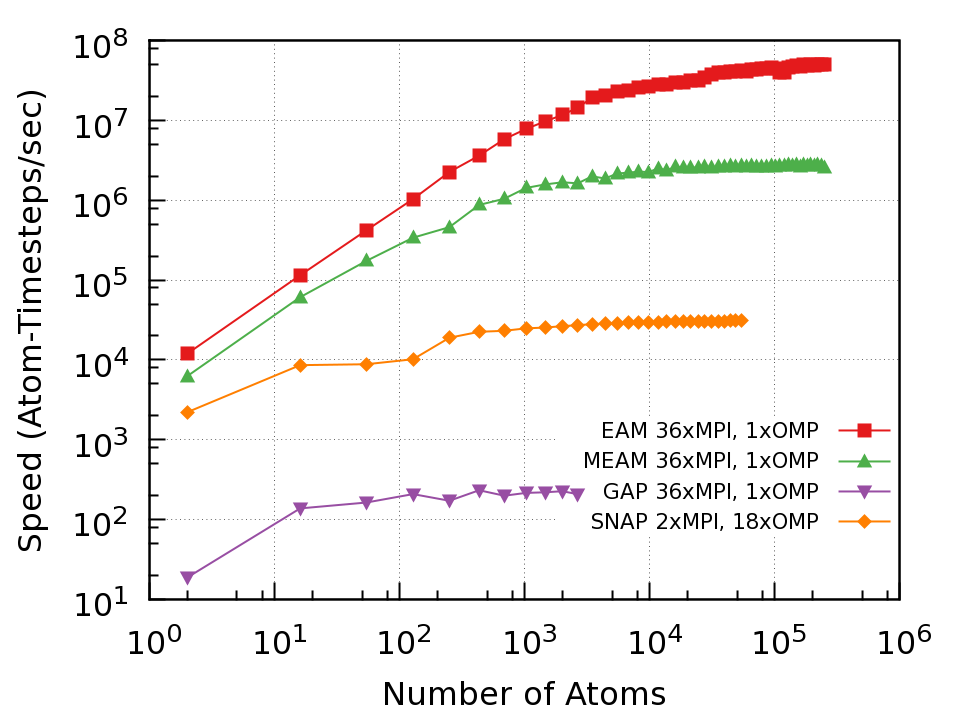} 
\caption{\label{fig3} Log-log plot of computational speed versus atom count for the different tungsten potentials using a single Intel Broadwell node. For each potential, the speed in atom-timesteps/sec becomes roughly constant for sufficiently large atom count, reflecting the linear scaling of the underlying LAMMPS neighborlist algorithm.}
\end{figure}

We measure computational cost by running a short tungsten MD simulation with each potential on one node of a high end computer. We repeat
this over a wide range of atom counts, because the optimal number of atoms per node is different for different potentials.
For each potential, we experimented with the different available build options and show results for the best choice 
for each potential. 
We used a single Intel Broadwell node of Sandia's Serrano computing cluster.  
The node consists of two, 2.1 GHz Intel Broadwell E5-2695 v4 processors with 18 physical cores each.
At runtime, the performance of each potential was optimized by varying the number of MPI tasks while
adjusting the number of OpenMP threads per MPI task to keep the total number of threads on the
node equal to 36, the total number of cores. 
We did not enable hyperthreading, non-standard NUMA settings, or leaving some core idle.  
The timings for each potential are shown in Figure \ref{fig3}.
Several general trends are manifested and are worth discussing here. 
Each potential undergoes a transition from a roughly constant maximum speed to strongly decreasing speed as the problem size is reduced.  
The maximum speed decreases by six orders of magnitude in 
going from EAM to MEAM to SNAP to GAP.  The critical atom count is smaller for the more expensive potentials.

%%%
%%%

We conclude that the more expensive potentials, specifically SNAP and GAP, scale much better down to small atom counts per node.
This can be seen by the magnitude of the fall off at low atom counts as compared to the performance at high atom counts each potential.
For EAM, there there is approximately a $5\times10^{3}$ difference in performance in going from the largest to the smallest atom counts, 
while for SNAP there is only a factor of ten difference.
In a practical sense, this means that a fixed-size SNAP MD simulation can be distributed across a much larger
 compute platform than would be possible with the same fixed-size EAM or MEAM simulation. Consequently,
 the difference in the time to solution between SNAP and EAM will be much less than the raw computational cost would indicate.  
 This is an important aspect of high performance computing to consider as the push toward exascale machines comes to fruition in the next few years.
Efficiently utilizing all of the resources on extreme-scale platforms is a challenging problem,
which can be mitigated by deployment of high-accuracy compute-intensive MD potentials such as SNAP. 

\section{\label{sec:level1}Conclusions}
In this report we have reported on the development of a quantum accurate SNAP potential for tungsten as well as reported data showing accurate helium defect properties within the tungsten matrix by merging the SNAP IAP with the repulsive pairwise W-He interactions from Juslin \cite{Juslin2013}. 
We believe that the improved accuracy for both W and W-He defects will play an important role in the prediction of He bubble formation and subsequent failure mechanisms of plasma-facing components.

\begin{acknowledgments}
We would like to thank Brian Wirth, Art Voter and Steve Plimpton for fruitful discussions throughout the work presented here.
Sandia National Laboratories is a multi-program laboratory managed and operated by Sandia Corporation, a wholly owned subsidiary of Lockheed Martin Corporation, for the US Department of Energy National Nuclear Security Administration under Contract No. DE- AC04-94AL85000.
\end{acknowledgments}

\end{document}